# Influence of random roughness on the Casimir force at small separations

P.J. van Zwol, G. Palasantzas[*] and J. Th. M. De Hosson

Department of Applied Physics, Netherlands Institute for Metals Research and Zernike Institute for Advanced Materials, University of Groningen, Nijenborgh 4, 9747 AG Groningen, the Netherlands.

**Abstract**

The influence of random surface roughness of Au films on the Casimir force is explored with atomic force microscopy in the plate-sphere geometry. The experimental results are compared to theoretical predictions for separations ranging between *20* and *200 nm*. The optical response and roughness of the Au films were measured and used as input in theoretical predictions. It is found that at separations below *100 nm,* the roughness effect is manifested through a strong deviation from the normal scaling of the force with separation distance. Moreover, deviations from theoretical predictions based on perturbation theory can be larger than *100%*.

PACS numbers: 68.55.Jk, 68.37.Ps, 85.85.+j, 78.68.+m

---

[*]Corresponding author: g.palasantzas@rug.nl

When the proximity between material objects becomes of the order of nanometers up to a few microns, a regime is entered where forces quantum mechanical in nature, namely, *van der Waals* and *Casimir* forces, become operative [1]. Historically, the Casimir force has been considered to result from the perturbation of zero point vacuum fluctuations by conducting plates [1]. Because of its relatively short range, the Casimir force is now starting to take on technological importance in the operation of micro/nanoelectromechanical systems (MEMS/NEMS) at separations *<200 nm*, e.g., micro oscillator devices, micro/nano switches, nanoscale tweezers or actuators [2-9]. Moreover, from a fundamental point of view the Casimir force plays important role in the search of hypothetical forces beyond the standard model [6]. The early force measurements by Sparnaay and van Blockland and Overbeek [1] gave evidense on the existence of the Casimir force. Higher accuracy measurements by Lamoreaux with the use of torsion pendulum [7] initiated further detailed investigations of the Casimir force. It was also measured accurately by other groups in the plate-sphere setup with the Atomic Force Microscope (AFM), and micro oscillator devices [8, 9]. Other geometries were also investigated, e.g., crossed cylinders [10], and parallel plates [11].

So far most measurements of the Casimir force were performed in the plate-sphere geometry to avoid problems with the plate-plate alignment. The Proximity Force Approximation (PFA), which is accurate for small plate-sphere separations $D$ compared to the sphere radius $R_{sph}$ ($D/R_{sph} <0.01$) [12], was then used to compare experiments and predictions based on theory. Nonetheless, the experimental situation is far from ideal as one does not deal with perfectly conducting flat mirrors, which otherwise lead for the plate-sphere geometry to the material independent force $F_{cas} \approx (\pi^3/360) R_{sph} (\hbar c/D^3)$ with $\hbar$ the Planck constant and $c$ the velocity of light [1, 8]. Indeed, the deposited metal coatings for both substrates and spheres are not perfect reflectors and have rough surfaces. The problem becomes clear if one realizes that variations in the available optical data [13] could lead to variations in the Casimir force of up to *10 %* [14].

Moreover, surface roughness (of sphere and/or plate) can be a formidable barrier in comparing experiment with predictions derived from perturbation theory (incorporating both roughness and finite conductivity corrections) [15] for separations *<100 nm*.

Therefore, a quest exists for proper investigations of the influence of roughness if one wishes to compare the Casimir effect theory with experiment for real materials at separations below *100 nm*. However, up to now a systematic study of controlled surface roughness growth and its influence on the Casimir force, in combination with direct optical characterization of the deposited metal films, is still lacking. This will be the topic of the present work, where we will also compare force measurements to predictions from the perturbative scattering theory, which incorporates roughness corrections in terms of the complete surface roughness spectrum [15].

Polysterene spheres with a diameter of *100 μm* and a 1.4%-deviation from sphericity (the diameter was measured with a Scanning Electron Microscope), were glued on a *450 μm* long Au coated cantilever. A relatively stiff cantilever is used to reduce the jump to contact, cantilever bending due to Au evaporation, and errors in the deflection correction. AFM was used to determine the sphere roughness (prior to Au coating), which gave a *1.2 nm* RMS roughness amplitude over an area of *25 μm$^2$*. Further, the spheres were plasma sputtered for electrical contact with the cantilever, and then were coated with *100 nm* Au at a rate of *0.6 nm/sec* in an evaporator kept at a base pressure of *10$^{-6}$ mbar*. Si-oxide wafers were used as substrates and coated by Au layers in the thicknesses between *100* and *1600 nm*, and under identical growth conditions with the Au coating on the sphere (see Fig. 1).

From five topography AFM scans on different locations per substrate/sphere surface area (Fig. 1), we measured the height-height correlation function $g(r) = <[h(r) - h(0)]^2>$ for the roughness analysis. *h(r)* is the height fluctuation, which is assumed to be a single valued function of the in-plane postion *r=(x,y)* [17, 18], and *<...>* indicates ensemble average over five scans on different surface locations. In many cases, the non equilibrium growth of random rough surfaces, as the Au films in this study, lead to the so called self-affine scaling [16-18]. In this case, the roughness is characterized by the RMS roughness amplitude *w*, the lateral roughness correlation length $\xi$, and the roughness exponent *H (0<H<1)* that characterize the degree of surface irregularity at short length scales ($<\xi$). After Au deposition, the roughness measurements for the sphere yielded $w_{sph}$=4.7 nm, $\xi_{sph}$=33 nm, and $H_{sph}$=0.8±0.05. Figure 2 shows the evolution of the roughness parameters for the Au substrates as a function of film thickness. The roughness exponent was constant *H=0.9±0.05* in agreement with former growth studies of Au films [17, 18]. The obtained roughness parameters were used as input to the perturbation theory [15, 18] to calculate the Casimir force.

Furthermore, the Picoforce AFM from Veeco [19] was used for the Casimir force measurements (Fig. 3) following the procedure outlined in [8]. Moreover, averaging over *30* force curves was used when the Casimir force was measured [19] The plate-sphere separation $D=D_{piezo}+d_0-d_{defl}$ was measured with respect to the point of contact with the surface, where $D_{piezo}$ is the piezo movement, $d_0$ is the distance on contact due to substrate and sphere roughness, and $d_{defl}=mF_{pd}$ is the cantilever deflection correction. $F_{pd}$ is the photodiode difference signal and *m* the deflection coefficient, i.e., rate of change of separation per unit photodiode difference signal [8]. The cantilever spring constant *k* was determined electrostatically with an error of *2 %* [21, 22], and it was recalibrated for all films without any measurable roughness effect on the electrostatic force. Indeed, the electrostatic force between the large sphere and the flat surface is given by [8] $F_{el} \approx [(V_1-V_c)^2/2]\sum_{n=1}^{+\infty} \csc h(na)(\coth(a)-n\coth(na))$ with $V_1$ the applied voltage on the plate, $V_c$ the residual or contact potential on the grounded sphere, and $a = \cosh^{-1}(1+D/R_{sph})$. The contact potential $V_c$ was determined electrostatically, and it was found to be $V_c \leq 25$ mV

*(*error $\leq 5$ mV*)* [8, 21]. Once $k$ and $V_c$ are known, $d_0$ is determined electrostatically (upper inset in Fig. 3) [8, 23]. Besides the electrostatic force and the Casimir force, a roughly linear signal was superimposed on the curves due to light reflected on the substrate and picked up by the AFM photodiode. This linear signal and the Casimir force have to be removed from the electrostatic calibration curves. The linear signal is filtered out by fitting at large separations (>*300 nm*) for the Casimir force curves [8]. Our measurements were restricted to separations below *200 nm* where the Casimir force is much larger than the linear signal. The small separation limit is restricted by the jump to a contact and surface roughness.

The error in the measured force can be estimated as follows. The error due to sphere nonuniformity on the force is ~*1.4 %* (since $F_{cas}$~$R_{sph}$), and from the $k$-constant, which is used to translate cantilever deflection to force, is ~*2 %*. Since $k=0.235$ *N/m*, the *2 %* variation found for the deflection coefficient *m* will give only a *0.2 nm* error at the closest separation and thus a relative error in the force <*2 %* [20]. The contact potential $V_c$ ($\leq 25$ *mV*) gives an error for the force smaller than *1%* at the separations <*200 nm*, while the error in the separation upon contact $d_0$ due to roughness is a predominant factor at separations < *100 nm*. Indeed, the measured error in $d_0$ was $\Delta d_o$ ~*1 nm*, leading to a relative error in the Casimir force of *2-15%* depending on separation [20]. We should stress, however, that $d_0$ corresponds to the peak distribution of the roughness and varies consistently with it. In fact, $d_0$ was found to vary almost linearly with the sum of the RMS roughness amplitudes of sphere ($w_{sph}$) and substrate ($w$): $d_0 \approx c(w+w_{sph})$ with $c=3.7\pm0.3$ (see upper inset, Fig. 3). Moreover, we accounted sufficiently well for any vertical drift since $\Delta d_o$ ~*1 nm* as determined by electrostatic measurements, which indicates that the AFM is stable within *1 nm*. The thermal drift was corrected because all measured Casimir and electrostatic force curves were shifted to have the same point of contact with the surface. Finally, for separations *D>100 nm* thermal and external vibrations become significant, leading to *2-5 pN* or up to *30%* error in the force at *200 nm*. Therefore, we can infer that the error in the force measurement for *D<200 nm* is on average *~10-15 %* [24].

Finally, since the error analysis in the Casimir force measurements is complete, we will discuss the force curves in comparison to theoretical predictions. In order to gauge the error in the theoretical force prediction due to variations of the optical response of the films, the latter was measured by ellipsometry for wavelengths *137 nm - 33 μm* [25] (Fig. 4). Note that the Au films under consideration are much thicker (*≥100 nm*) than the skin depth, and therefore are optically opaque [14]. The response at lower wavelengths has only a marginal effect on the Casimir force, and it was taken from Palik's handbook data [13]. For wavelengths *>33 μm* the Drude model was used for optical response. The Drude parameters were found by fitting the data in the infrared range [25]. The fit yielded the plasma frequency values $\omega_p$=*7.2, 7.5 and 8.2±0.2eV,* and relaxation frequency $\omega_t$=*0.055, 0.057* and *0.065 ± 0.005 eV*, for the *400, 200* and *100 nm* films respectively. This can be compared with fitting Palik's data yielding *7.5 eV* and *0.061 eV* [26]. This is lower than the theoretical value for perfect Au films with $\omega_p$=*9 eV* since our samples were not annealed and may contain voids and grains. Note that when fitting Palik's data with the Drude parameters for perfect Au coatings (with fixed $\omega_p$=*9eV* a value for $\omega_t$=*0.035 eV* is obtained), the error can be more than *10%* in the Casimir force calculation [14]. Upon substitution of the measured optical data into the theory a variation of less than *10 %* was found in the Casimir force for the different Au coatings. However, this is smaller in magnitude than the average error of *~10-15 %* due to roughness at separations above *50 nm*.

Furthermore, Fig. 3 shows the scattering theory predictions [15] for *100 nm* and *1600 nm* thick films. Briefly, the Casimir force is given by $F=2\pi R_{sph} E_{PP}$ with $E_{PP}$ the Casimir energy in the plane-plane geometry. For weak roughness, $E_{PP}$ is given by [15]
$E_{PP} \cong E_{PP,flat} + (1/2)(\partial^2 E_{PP,flat}/\partial D^2) \sum_{m=1}^{2} [d^2q/(2\pi)^2] p_m(q) <|h_m(q)|^2>$ with $E_{PP,flat}$ the energy for flat surfaces given by the Lifshitz formula (which allows the incorporation of real optical data for the dielectric function of the Au films; see Fig. 4). $<|h_{mm}(q)|^2>$ are the roughness spectra (Fourier transform of g(r)) for the sphere ($m=1$) and plate ($m=2$) surfaces [18]. $P_m(q)$ is a response function related to photon scattering between the plates, which is an improvement of that of ref. [15] by incorporating into the formalism the measured dielectric function of the Au films. Note that, since our measurements took place at $T \approx 300\ K$ the finite temperature corrections to the force are ~1 % for $D<200\ nm$, and therefore are neglected [8]. The perturbative scattering theory and experiment are in agreement within the error of measurement ~10-15 % for separations $D>60\ nm$ (see also lower inset in Fig. 3). At smaller separations ($D<40\ nm$), the Casimir force is highly sensitive to optical characteristics of the films as will be explained in detail in a future publication [26].

For the thicker films, the measurement clearly shows the systematic deviation from the normal Casimir power law scaling. The deviation correlates to the evolution of the roughness amplitude and local surface slope with film thickness in Fig. 2. In fact, the pertubative theory [15] is valid for significant separations $D>>w+w_{sph}$ and weak local slopes $|\nabla h|<<1$ or equivalently $\rho_{rms}=<|\nabla h|^2>^{1/2}<<1$. The inset in Fig. 2 shows that $\rho_{rms}$ increases with increasing thickness, which indicates surface roughening [18]. Although $\rho_{rms}<1$, it still remains significant for the rougher films. Therefore, for separations $D<60\ nm$ where the roughness influence is significant, both conditions for the applicability of perturbative scattering theory [15] are violated (with $D$ comparable $w+w_{sph}$ being the dominant source of discrepancy for rougher films). Notably, in the regime of interest ($D<60\ nm$), e.g. for the 1600 nm thick film (Fig. 3), the second order scattering corrections can be as large as 30 %, which is an additional indicative factor for the strong influence of surface roughness.

The magnitude of the measured roughness effect on the Casimir force is of the order of *100 %* for thicker films (which are also rougher films) at short separations *(< 80 nm)*, while at larger separations the scaling law is recovered and agreement with theory is restored. Qualitatively the roughness effect could be reproduced (for illustrative purposes) by performing a direct integration using the Lifshitz formula to compute the Force between rough surfaces by point to point (using the AFM topography scans) summation (non perturbative PFA) and average over five measured roughness scans (Fig. 5) [12]. Although the non perturbative PFA approach is qualitative, it can be used to obtain an estimate of the force at extreme close proximity (*~2 nm* above the point upon contact), where the roughness has an enormous influence on the Casimir force. This explains the jump to contact only partially, since contributions due to local capillary forces around surface protrusions will play role [27].

In conclusion, we have shown that moderate roughness significantly alters the Casimir force at separations below *100 nm* following the roughness evolution. Since the measured Casimir forces on six different samples coincide at large separations, within the measured error, the measurements are reproducible. In addition, they have been reproduced with independent measurements, using different cantilevers. Furthermore, the contact distance $d_0$ corresponds to the peak distribution of the surface roughness making our force measurements also conclusive at small separations. The effect of roughness manifests itself through a strong deviation from the normal scaling of the force in the plate-sphere geometry, leading to a deviation from theoretical predictions by more than *100%*.

**Acknowledgements**

The research was carried out under project number MC3.05242 in the framework of the Strategic Research programme of the Netherlands Institute for Metals Research (NIMR). Financial support from the NIMR is gratefully acknowledged. We acknowledge useful discussions with A. Lambrecht, S. Reynaud, S. K. Lamoreaux, V. B. Svetovoy, A. A. Maradudin, C. Capasso, J. Munday, C. Binns, and D. Iannuzzi.

**References**

[1] H. B. G. Casimir, Proc. K. Ned. Akad. Wet. 51, 793 (1948). For initial measurements of the Casimir effect see: M. J. Sparnaay, Physica (Utrecht) 24, 751 (1958); P. H. G. M. van Blockland and J. T. G. Overbeek, J. Chem. Soc. Faraday Trans. 74, 2637 (1978).

[2] K. L. Ekinci and M. L. Roukes, Rev. Sci. Instrum. 76, 061101 (2005)

[3] A. Cleland, Foundations of Nanomechanics (Springer, New York, 2003)

[4] D. Iannuzzi, M. Lisanti, F. Capasso, PNAS 102, 11989 (2005)

[5] F. M. Serry, D. Walliser, and G. J. Maclay, J. Appl. Phys. 84, 2501 (1997); [3] Wen-Hui Lin, Ya-Pu Zhao, Chaos, Solitons and Fractals 23, 1777 (2005); G. Palasantzas and J. Th. M. De Hosson, Phys. Rev. B 72, 115426 (2005); G. Palasantzas and J. Th. M. De Hosson, Phys. Rev. B 72, 121409 (2005); G. Palasantzas and J. Th. M. De Hosson, Surf. Sci. 600, 1450 (2006).

[6] R. Onofrio, New J. Phys. 8, 237 (2006)

[7] S. K. Lamoreaux, Phys. Rev. Lett, 78, 5 (1997).

[8] B. W. Harris, F. Chen, U. Mohideen, Phys. Rev. A. 62, 052109 (2000); M. Bordag, U. Mohideen, V. M. Mostepanenko, Phys. Rep. 353 (2001); F. Chen, G. L. Klimchitskaya, U. Mohideen, and V. M. Mostepanenko, Phys. Rev. A **69**, 022117 (2004); F. Chen and U. Mohideen, G. L. Klimchitskaya, V. M. Mostepanenko, Phys. Rev. A 74, 022103, 2006.

[9] H. B. Chan, V. A. Aksyuk, R. N. Kleiman, D. J. Bishop, F. Capasso, Science 291, 1941 (2001); R. Decca, E. Fischbach, G. L. Klimchitskaya, D. E. Krause, D. Ló´pez, and V. M. Mostepanenko, Phys. Rev. D 68, 116003 (2003); R.S. Decca, D. López, E. Fischbach, G.L. Klimchitskaya, D.E. Krause, and V.M. Mostepanenko, Ann. Phys. (N.Y.) 318, 37 (2005).

[10] T. Ederth, Phys. Rev. A, 62, 062104 (2000)

[11] G. Bressi, G. Carugno, R. Onofrio, G. Ruoso, Phys. Rev. Lett. 88 041804 (2002)

[12] H. Gies and K. Klingmuller, Phys. Rev. D, 74, 045002 (2006)

[13] E. D. Palik, Handbook of optical constants of Solids (Academic Press, 1995)

[14] I. Pirozhenko, A. Lambrecht, V. B. Svetovoy, New J. Phys. 8, 238 (2006)

[15] P. A. Maia Neto, A. Lambrecht and S. Reynaud, Phys. Rev A 72, 012115 (2005); P. A. Maia Neto, A. Lambrecht, S. Reynoud, Europhys. Lett. 69, 924 (2005).

[16] P. Meakin Phys. Rep. 235 1991 (1994)

[17] J. Krim and G. Palasantzas, Int. J. of Mod. Phys. B 9, 599 (1995).

[18] G. Palasantzas, Phys. Rev. B 48, 14472 (1993); 49, 5785 (1994); G. Palasantzas and J. Krim, Phys. Rev. Lett. 73, 3564 (1994); G. Palasantzas, Phys.Rev.E. 56, 1254 (1997). For self-affine roughness $g(r)$ scales as: $g(r) \propto r^{2H}$ if $r<<\xi$, and $g(r) = 2w^2$ if $r>>\xi$. Thus, $H$ is determined by the slope ($=2H$) of the linear part of $g(r)$ in a log-log plot, and $\xi$ by the crossover between a linear fit and the saturation regime that gives also $w$.

[19] http://www.veeco.com/products/details.php?cat=1&sub=1&pid=192. Creep and hysteresis are minimized by the closed loop scanner (noise error *<0.5 nm* and *0.1%* nonlinearity). The force curves in Fig. 3 consist of *$10^3$* points (averaged above *100 nm* if the difference in force between successive points was ≤*1 %*). Each force curve is measured in *0.5 sec* to minimize thermal drift.

[20] In Fig. 3, the actual theory plots show a close linear relation for the force vs. separation D as $F_{cas} \propto D^{-2.5}$, and therefore we obtain the relative error in the force $\Delta F_{cas}/F_{cas} \cong (2.5 \Delta D / D)$ with $\Delta D$ the error in separation. Thus, the error in $d_o$ is an experimental error and it will be transferred as an error in the force curve. In fact, upon repeating the force measurements one can see a variation in the force curves due to this error in $d_o$ (this point is also clarified in the work D. Iannuzzi, I. Gelfand, M. Lisanti, and F. Capasso (*arXiv:quant-ph/0312043, 2003*): *http://arxiv.org/PS_cache/quant-ph/pdf/0312/0312043v1.pdf*).

[21] For the electrostatic calibration we used *8* electrostatic force curves with $V_1$ between *±3 V* to *±4.5 V* (error *1 mV*) at the point of contact with the surface. The deflection sensitivity was (120±4) nm/V. The deflection correction $d_{defl}$ was applied to these curves, and their point of contact was shifted to the distance upon contact $d_0$. From the same curves we also determined $k$ and $V_c$ (at *3 μm* above the plate [8]).

[22] An error of *30 nm* in $d_0$ will result in *1%* error in $k$ since $F_{el} \sim 1/D$ in the plate-sphere setup.

[23] $d_o$ was determined from *60* curves with *6* different applied voltages between *250–500 mV*.

[24] S. K. Lamoreaux, Phys. Today, Feb. 2007 (p.40).

[25] H. G. Tompkins, W.A. McGahan, "*Spectroscopic Ellipsometry and Reflectometry*" (New York, NY: John Wiley & Sons, 1999).

[26] Extensive analysis of the optical data will be presented in the future (in preparation).

[27] Capillary condensation occurs for separation closer than about twice the Kelvin condensation length *(~1.5 nm* for water at room temperature and *50 %* relative humidity): R. Maboudian and R. T. Howe J. Vac. Sci. Technol. B 15, 1 (1997).

**Figure Captions**

**Figure 1**: Surface scans of all films and the sphere surface, the scan size is *500 nm* and the color range corresponds to indicated range by the color bar.

**Figure 2** Plots of the roughness amplitude (open squares) and lateral correlation length (filled dotts) vs. film thickness. The inset shows the local surface slope vs. film thickness.

**Figure 3 (a)** Casimir force measurements for different rough surfaces on a log-log scale for the various film thicknesses. *100 nm*: ◊, *200 nm:* *, *400 nm*: o, *800 nm*: Δ, and *1600 nm*: ∇. The perturbation theory calculations are shown for the *100 nm* film (dashes) and the *1600 nm* film (solid line). The upper inset shows the $d_0$ vs. $w^{\#}=w+w_{sph}$ relation, where a linear fit gives $c=3.7\pm0.3$. The inset below shows the relative error between theory- experiment for the *100 nm* and *1600 nm* film (in percent) with respect to the corresponding measured force.

**Figure 4**: Dielectric function at imaginary frequencies for Paliks data extended to the low frequency domain with Drude parameters *9 eV* and *34 meV*, and our films. The inset shows the imaginary part of the dielectric function from the raw data.

**Figure 5**: Numerical calculations of the roughness influence using the PFA and direct integration over the roughness scans down to *2-3 nm* above separation upon contact.

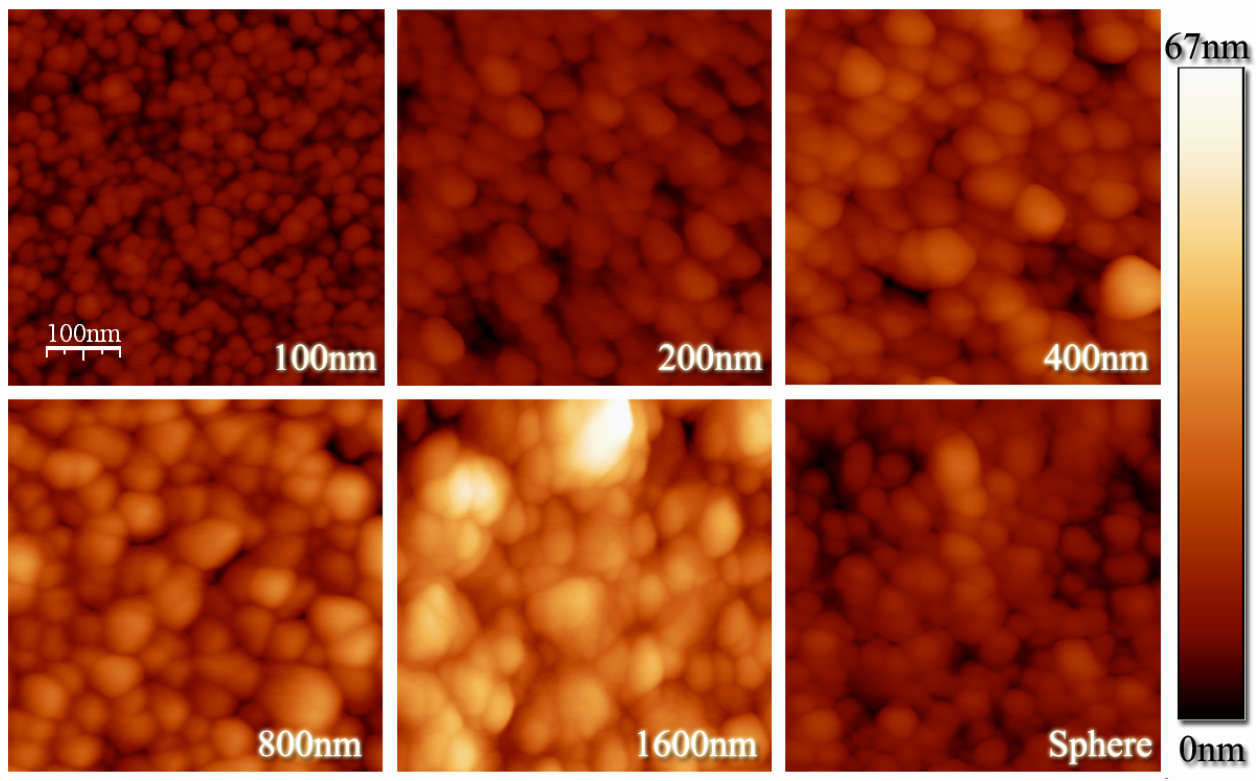

**Figure 1**

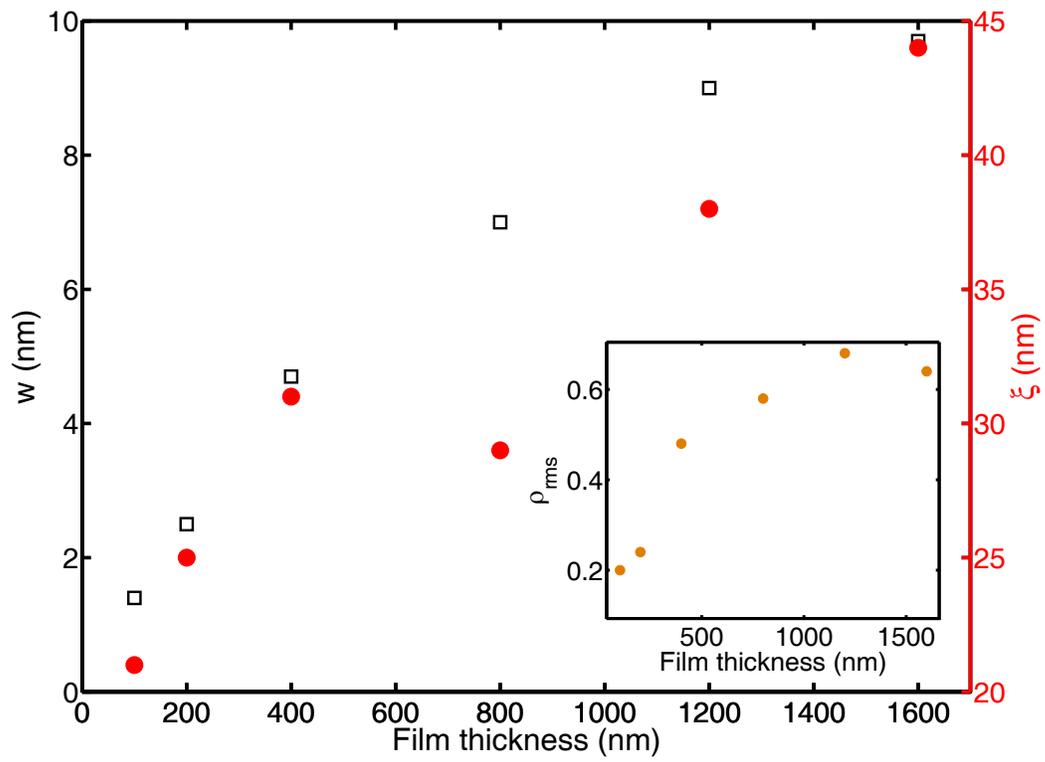

**Figure 2**

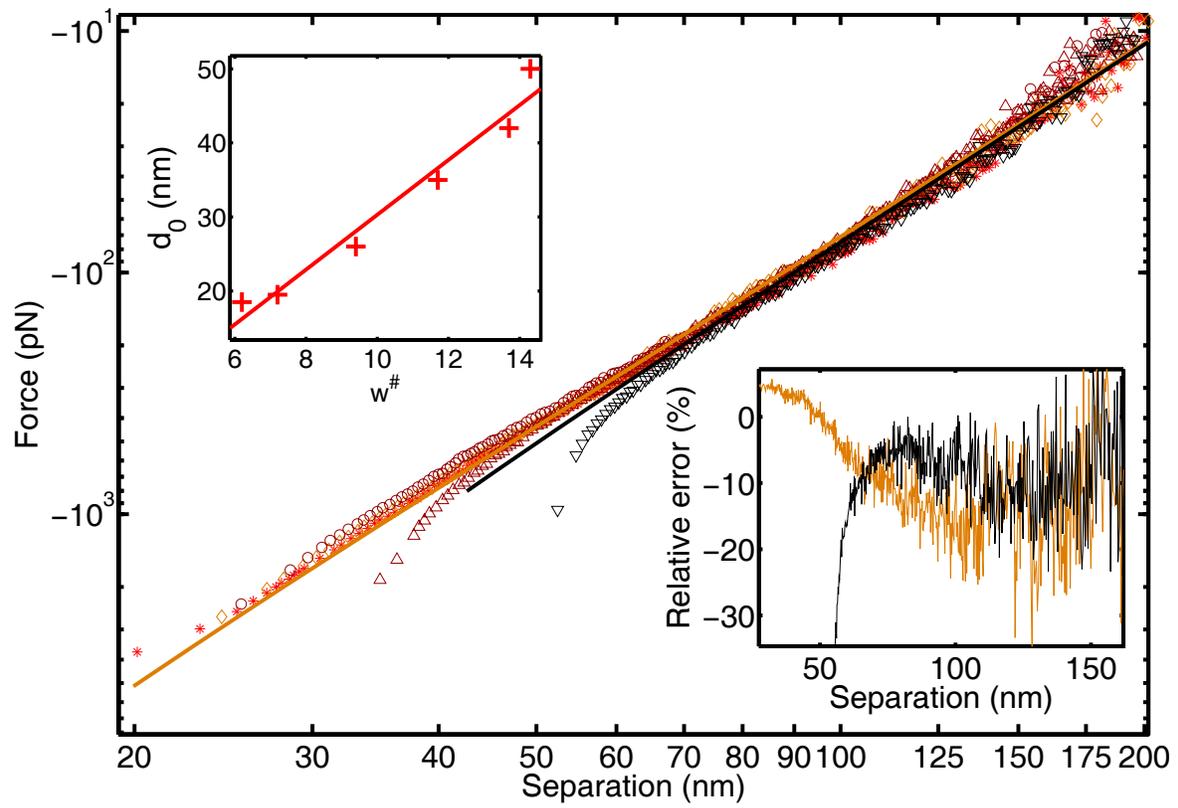

**Figure 3**

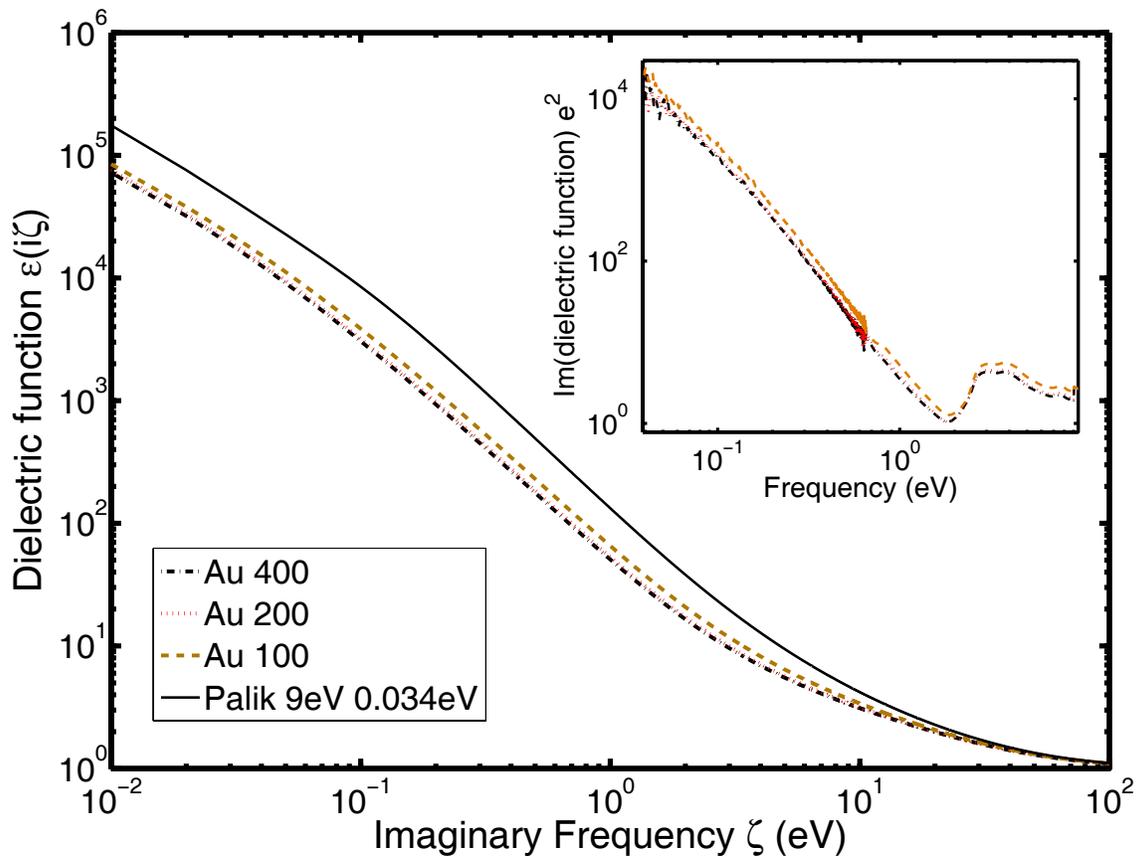

**Figure 4**

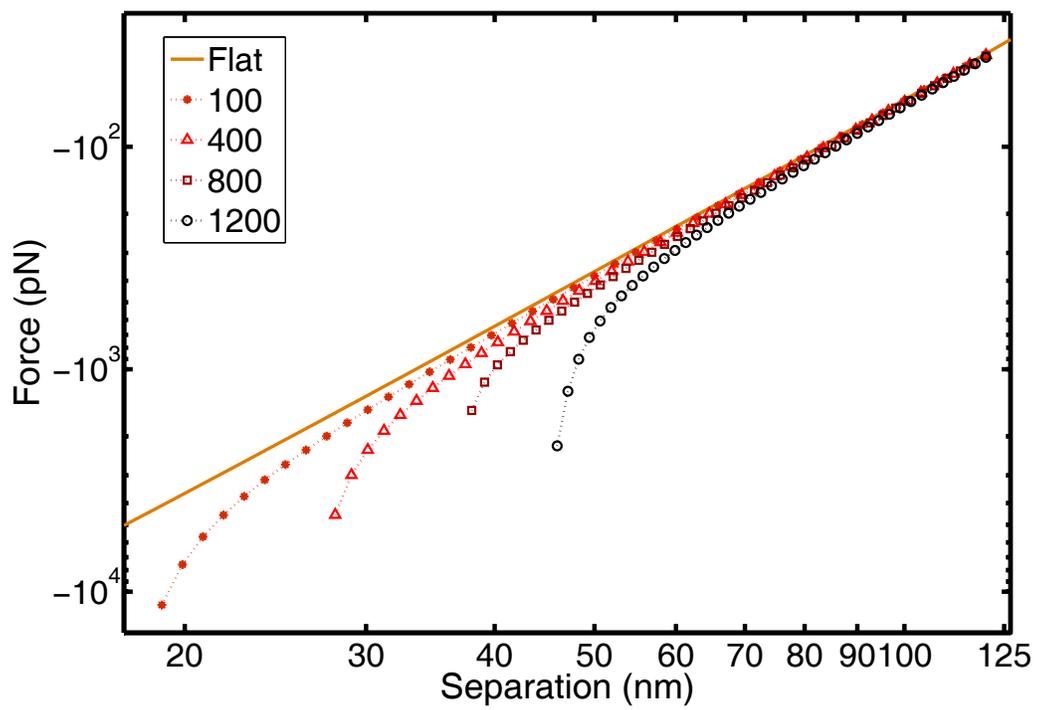

**Figure 5**